\documentclass[10pt,aps,prd,reprint,nofootinbib,floats,floatfix,amsfonts,amssymb,amsmath,preprintnumbers,notitlepage,superscriptaddress]{revtex4-1}
\usepackage{aastex_macros}
\usepackage{float}
\usepackage{graphicx}
\usepackage{caption}
\usepackage{subcaption}
\usepackage[utf8]{inputenc}
\usepackage[british]{babel}
\usepackage{isomath}
\usepackage[protrusion=true,tracking=true,kerning=true,spacing=true,final,babel=true]{microtype}
\usepackage{physics}
\usepackage{amsmath}
\usepackage{xcolor}
\usepackage{url}
\usepackage[final]{hyperref}

\begin{document}

\begin{flushright}
KCL-PH-TH-2022-07
\end{flushright}

\title{Detectability of the gravitational-wave background\texorpdfstring{\\}{} produced by magnetar giant flares}
\author{Nikolaos Kouvatsos}
\affiliation{Theoretical Particle Physics and Cosmology Group, \, Physics \, Department, \\ King's College London, \, University \, of London, \, Strand, \, London \, WC2R \, 2LS, \, UK}
\author{Paul D. Lasky}
\affiliation{School of Physics and Astronomy, Monash University, VIC 3800, Australia}
\affiliation{OzGrav: The ARC Centre of Excellence for Gravitational-wave Discovery, Clayton, VIC 3800, Australia}
\author{Ryan Quitzow-James}
\affiliation{Institute of Multi-messenger Astrophysics and Cosmology, Missouri University of Science and Technology, Rolla, MO 65409, USA}
\author{Mairi Sakellariadou}
\affiliation{Theoretical Particle Physics and Cosmology Group, \, Physics \, Department, \\ King's College London, \, University \, of London, \, Strand, \, London \, WC2R \, 2LS, \, UK}

\date{\today}

\begin{abstract}
We study the gravitational-wave background produced by f-mode oscillations of neutron stars triggered by magnetar giant flares. For the gravitational-wave energy, we use analytic formulae obtained via general relativistic magnetohydrodynamic simulations of strongly magnetized neutron stars. Assuming the magnetar giant flare rate is proportional to the star-formation rate, we show the gravitational-wave signal is likely undetectable by third-generation detectors such as the Einstein Telescope and Cosmic Explorer. We calculate the minimum value of the magnetic field and the magnetar giant flare rate necessary for such a signal to be detectable, and discuss these in the context of our current understanding of magnetar flares throughout the Universe.
\end{abstract}

\maketitle

\section{INTRODUCTION}
Magnetar giant flares (MGFs) are extremely luminous transients arising from violent explosions in the magnetospheres of strongly magnetised neutron stars known as magnetars. Several such flares have been detected thus far, with tails typically lasting $\sim (200-400)$ s and unleashing $\gtrsim10^{44}$ erg of energy in hard X rays and soft gamma rays \cite{10.1111/j.1365-2966.2010.17038.x,Israel_2005,2008}. While this energy is emitted as electromagnetic radiation, MGFs may also lead to the emission of gravitational waves (GWs) via the excitation of oscillation modes inside the neutron stars \cite{1999_a,10.1111/j.1365-2966.2011.19515.x}.
%This has motivated several theoretical and observational studies on the detection of GWs resulting from MGFs \cite{PhysRevD.83.104014,PhysRevD.85.024030,2008_a,2011_a,2019_p}.
This has motivated several theoretical and observational studies on the detection of GWs resulting from MGFs \cite[e.g.,][]{Matone_2007,PhysRevD.76.062003,Kalmus_2007,2008_a,Abbott_2008,2009PhDT.......272K,Abadie_2012,PhysRevD.87.103008,Macquet_2021,PhysRevD.83.104014,PhysRevD.85.024030}.

Theoretical studies of GWs from MGFs have primarily focused on the f~mode \cite[e.g.,][]{Ioka2001,PhysRevD.83.104014,10.1111/j.1365-2966.2011.19515.x,PhysRevD.85.024030,lasky2012gravitational,2012_l}, which is considered the most efficient emission mode~\cite{Detweiler1975, McDermott1988}. Some of these works~\cite{PhysRevD.85.024030,2012_l} performed general relativistic magnetohydromagnetic simulations assuming different equations of state for polytropic neutron stars showing that the f~mode may just be detectable from a Galactic MGF with third-generation gravitational-wave observatories such as the Einstein Telescope (ET)~\cite{EinsteinTelescope} or Cosmic Explorer (CE)~\cite{CosmicExplorer}. Other oscillation modes, such as Alfv\'en modes and g~modes, may also produce lower frequency GWs, but their contribution is likely small compared to the dominant f~mode \cite{WATTS20071446,10.1111/j.1365-2966.2010.18009.x} (although see~\cite{PhysRevD.85.024030}).
%\cite{lasky2012gravitational} perform numerical relativity simulations for a poloidal magnetic field, considering various equations of state and magnetic field strengths, to obtain an analytic expression for the GW strain and energy.
%No evidence of GWs from a single MGF has been found to date \cite{2008_a,2011_a,2019_p}.
%No evidence of GWs from magnetar short bursts \cite{Kalmus_2007,2008_a,PhysRevD.80.042001,Abbott_2009,2011_a,quitzowjames2016thesis,Quitzow_James_2017,2019_p,schale2019search} or MGFs \cite{Matone_2007,PhysRevD.76.062003,Kalmus_2007,2008_a,Abbott_2008,2009PhDT.......272K,Abadie_2012,Macquet_2021}\footnote{While \cite{PhysRevLett.113.011102} was looking for GWs from GRBs, it was later determined one of the GRBs in the data, GRB 070222, was likely an extra-galactic MGF \cite{2021_3,Macquet_2021}.} has been found to date.
%No evidence of GWs from magnetar short bursts \cite{Kalmus_2007,2008_a,PhysRevD.80.042001,Abbott_2009,2011_a,quitzowjames2016thesis,Quitzow_James_2017,2019_p,schale2019search} or MGFs \cite{Matone_2007,PhysRevD.76.062003,Kalmus_2007,2008_a,Abbott_2008,2009PhDT.......272K,Abadie_2012,Macquet_2021}\footnote{GRB 070222 was part of the data analyzed by \cite{PhysRevLett.113.011102}, a search for GWs from gamma ray bursts (GRBs), and was later shown to probably be an extra-galactic MGF \cite{2021_3,Macquet_2021}.} has been found to date.
No evidence of GWs from magnetar short bursts \cite{Kalmus_2007,2008_a,PhysRevD.80.042001,Abbott_2009,2011_a,quitzowjames2016thesis,Quitzow_James_2017,2019_p,schale2019search} or MGFs \cite{Matone_2007,PhysRevD.76.062003,Kalmus_2007,2008_a,Abbott_2008,2009PhDT.......272K,Abadie_2012,Macquet_2021} has been found to date.

In this study, we explore the contribution of MGFs to the GW background. Previous studies of the GW background from magnetars have considered the emission of GWs from magnetars due to non-zero stellar ellipticities from internal magnetic fields~
\cite[e.g.,][]{refId0,2008_k,10.1111/j.1365-2966.2010.17861.x,PhysRevD.87.042002,universe7100381}. We follow the same main principles to construct the background, but for the GW energy we consider the excitation of the f~mode arising from MGFs using the results of \cite{lasky2012gravitational} for the GW strain. We obtain a normalised GW energy spectrum for the background from MGFs that we compare with the third-generation (3g) detector sensitivities.

The rest of the paper is organised as follows. In Section~\ref{II}, we outline our model used to calculate the GW background. In Section~\ref{III}, we obtain the GW energy spectrum of MGFs. Then, in Section \ref{IV}, we examine the detectability of the resulting signal assuming a network consisting of ET and CE and place constraints on the minimum required MGF rate and magnetic field strength for detection. We present our conclusions in Section~\ref{V}.

\section{GW BACKGROUND MODEL}\label{II}
%We follow closely the procedure outlined in \cite{regimbau_gravitational_2006,marassi_stochastic_2011,wu_accessibility_2013,chowdhury_stochastic_2021,phinney_practical_2001,finkel_stochastic_2021,regimbau_astrophysical_2008}. 
The GW background is described by the normalised GW energy spectrum \cite{PhysRevD.59.102001}
\begin{equation}
    \Omega_{\rm GW}(f_{\rm o})=\frac{1}{c^2\rho_{\rm c}}\frac{{\rm d}\rho_{\rm GW}}{{\rm d}\ln f}\bigg\vert_{f_{\rm o}},
    \label{1}
\end{equation}
where $\rho_{\rm c}=3H_0^2/(8\pi G)$ is the critical density, $H_0$ is Hubble's constant, $c$ the speed of light, $G$ Newton's gravitational constant, and $f_{\rm o}$ the emitted GW frequency as measured in the observer frame. For astrophysical sources, Eq.~($\ref{1}$) can be rewritten as \cite{2008_k}
\begin{equation}
    \Omega_{\rm GW}(f_{\rm o})=\frac{f_{\rm o}}{c^3\rho_{\rm c}}F(f_{\rm o}),
\end{equation}
where
\begin{equation}
    F(f_{\rm o})=\int_0^{z_{\rm max}}\frac{R_{\rm MGF}(z)}{4\pi d_c^2(z)}\frac{{\rm d}E_{\rm GW}}{{\rm d}f}\bigg\vert_{f_{\rm s}}dz,
\end{equation}
is the integrated flux density, $R_{\rm MGF}(z)$ is the MGF rate as a function of redshift $z$, $d_{\rm c}(z)$ the comoving distance, ${\rm d}E_{\rm GW}/{\rm d}f$ the GW energy spectrum, and $f_{\rm s}=f_{\rm o}(1+z)$ the GW frequency in the source frame.

The rate $R_{\rm MGF}$ can be expressed in terms of the MGF rate per unit comoving volume, $R_{{\rm MGF}(V)}$, as
\begin{equation}
    R_{\rm MGF}=\frac{R_{{\rm MGF}(V)}}{1+z}\frac{{\rm d}V}{{\rm d}z}=\frac{4\pi c}{H_0}\frac{R_{{\rm MGF}(V)}d_{\rm c}^2(z)}{(1+z)E(z)},
    \label{4}
\end{equation}
where $E(z)=\sqrt{\Omega_{\rm m}(1+z)^3+\Omega_\Lambda}$, with $\Omega_{\rm m}$ and $\Omega_{\Lambda}$ the energy density of baryonic matter and dark energy, respectively. We adopt a flat $\Lambda$CDM model with $\Omega_{\rm m}=0.311$, $\Omega_\Lambda=0.689$ \cite{2020} and $H_0=67.7 {\rm km}/{\rm s}/{\rm Mpc}$ \cite{2021_2}. The $(1+z)$ in the denominator, Eq.~($\ref{4}$), accounts for cosmic expansion.

We obtain the MGF rate assuming a linear scaling with the star formation rate (SFR). We note that there is a delay time between the birth of a star and its evolution into a magnetar. However, since the lifetime of a star (mass from $8M_{\odot}$ to $40M_{\odot}$ \cite{PhysRevD.87.042002}) turning into a neutron star is at most of order $10^7$ yr \cite{noauthor_stellar_nodate} and magnetar magnetic fields decay within $10^5$ yr \cite{mereghetti_pulsars_2013,Mondal_2021}, we deduce that adding a delay time has a negligible impact on our results.

%Through our analysis, the two parameters characterised by large uncertainty and therefore we vary are $\lambda$ and $B_{\rm pole}$.

We relate $R_{{\rm MGF}(V)}$ to the SFR via a proportionality constant $\lambda$:
\begin{equation}
    R_{{\rm MGF}(V)}(z)=\lambda R_*(z),
    \label{5}
\end{equation}
where
\begin{equation}
    R_*(z)=0.015\frac{(1+z)^{2.7}}{1+[(1+z)/2.9]^{5.6}}M_\odot {\rm yr}^{-1}{\rm Mpc}^{-3},
    \label{6}
\end{equation}
denotes the SFR, which is taken up to $z_{max}=8$ \cite{doi:10.1146/annurev-astro-081811-125615}.

We define $\lambda$ as the proportionality constant that relates the MGF rate to the SFR (we stress that this definition is different from the typical one that exists in the literature \cite{PhysRevD.87.042002,refId0}). It is equal to the mass fraction of stars that is converted into magnetars times the mean number of MGFs per magnetar. We presume that a magnetar will emit a number of MGFs during its lifetime, the value of which is not well-known. Assuming that all neutron stars are born as magnetars \cite{heras2012magnetar}, we assume an upper limit of $0.01M_{\odot}^{-1}$ for the mass fraction that is converted into magnetars. We then multiply this by the number of MGFs per magnetar to obtain $\lambda$. For example, a mean value of $10^2$ MGFs per magnetar yields $\lambda=1M_{\odot}^{-1}$, while a value of $10^4$ MGFs per magnetar results in $\lambda=10^2M_{\odot}^{-1}$. The authors of Ref.~\cite{2021_3} inferred a local (i.e., $z=0$) volumetric rate $R_{{\rm MGF}(V)}(0)=3.8\times10^5{\rm Gpc}^{-3}{\rm yr}^{-1}$. Plugging this value in Eq.~(\ref{5}) and solving for $z=0$ (today) yields  $\lambda=0.025 M_{\odot}^{-1}$ \cite{2021_3}. We use this as our reference value for $\lambda$, but also explore alternative values to understand the difference it makes on predictions for the stochastic GW background.

Our final expression for the normalised GW energy spectrum then reads:
\begin{align}
    \Omega_{\rm GW}(f_{\rm o})=&\frac{\lambda f_{\rm o}}{\rho_{\rm c}H_0c^2}\times\notag\\
    &\times\int_0^{z_{max}}\frac{R_*(z)\frac{{\rm d}E_{\rm GW}}{{\rm d}f}\Big\vert_{f_s}}{(1+z)\sqrt{\Omega_{\rm m}(1+z)^3+\Omega_\Lambda}}{\rm d}z.
    \label{7}
\end{align}
%and it remains to calculate the GW energy spectrum $\frac{{\rm d}E_{\rm GW}}{{\rm d}f}\Big\vert_{f_{\rm s}}$ of MGFs.

\section{GW ENERGY SPECTRUM}\label{III}
%Previous studies of the GW background of magnetars have modeled the emission of GW energy from a magnetar due to its fast rotation when the spin axis is misaligned with the magnetic distortion symmetry axis, assuming an initial angle of $\pi/2$
%\cite{refId0,2008_k,10.1111/j.1365-2966.2010.17861.x,PhysRevD.87.042002,universe7100381}. This results in a magnetic torque and a time-varying quadrupole moment.
%; ergo the emission of GWs. 
%Instead, we study individual bursts of GW energy as MGFs due to the excitation of the f-mode. 

General relativistic magnetohydrodynamic simulations of magnetars designed to mimic the internal magnetic field rearrangement immediately following an MGF show scaling relations between the star's magnetic field strength at the pole $B_{\rm pole}$ (assuming a poloidal configuration for the magnetic field), the emitted gravitational-wave strain in the f~mode, and the stellar mass $M$ and radius $R$ \cite{lasky2012gravitational}: 
\begin{align}
    h_{\rm max}=&8.5\times 10^{-28}\times\notag\\
    &\times\frac{10{\rm kpc}}{d}\bigg(\frac{R}{10{\rm km}}\bigg)^{4.8}\bigg(\frac{M}{M_{\odot}}\bigg)^{1.8}\bigg(\frac{B_{\rm pole}}{10^{15}{\rm G}}\bigg)^{2.9}.
    \label{8}
\end{align}
We model the GW strain as an exponentially decaying sinusoid peaked at the f-mode frequency:
\begin{equation}
    h=h_{\rm max}\sin(2\pi f_{\rm fmode})e^{-\frac{t}{\tau}},
    \label{9}
\end{equation}
where $f_{\rm fmode}$ is the f-mode frequency and $\tau$ is the decay constant. We consider a typical mass $M=1.4M_{\odot}$ and a radius $R=13$ km, consistent with \cite{Raaijmakers_2020,2021}, and a single value of $B_{\rm pole}$ to describe the whole magnetar population in the Universe. The corresponding equation of state yields  $f_{\rm fmode}=1883.1$ Hz and $\tau=0.25$ s, using \cite{1998}. 

\begin{figure}
\centering
\includegraphics[scale=0.65]{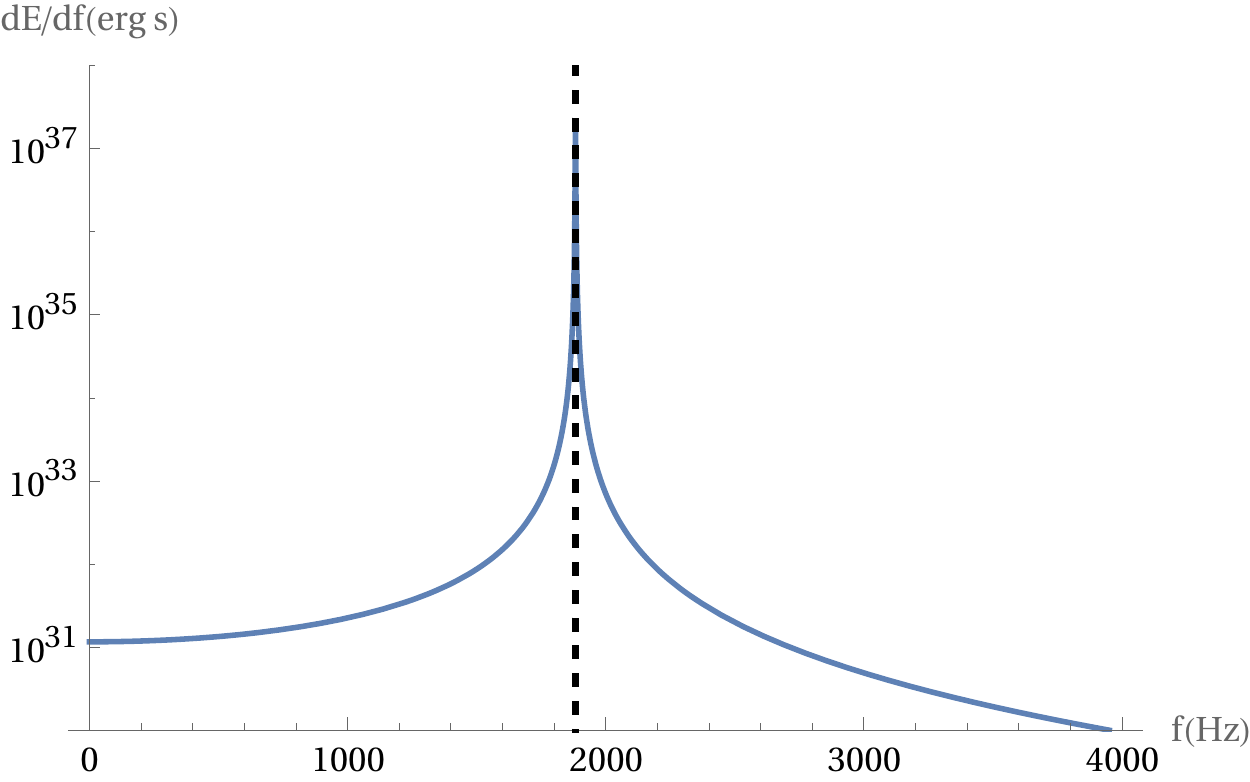}
\caption{The GW energy spectrum emitted via a single MGF from a magnetar with $B_{\rm pole}=10^{15}$ G. Treating the GW strain as an exponentially decaying sinusoid leads to a sharp peak at the $f_{\rm fmode}$ frequency.}
\label{dEdf}
\end{figure}

The total GW energy emitted is \cite{10.1111/j.1365-2966.2011.19515.x}
\begin{align}
    E_{\rm GW}=&\frac{2\pi^2d^2f^2_{\rm fmode}c^3}{G}\int_{-\infty}^{+\infty}\vert h(t)\vert^2{\rm d}t\notag\\
    =&\frac{2\pi^2d^2f^2_{\rm fmode}c^3}{G}\int_{-\infty}^{+\infty}\vert \hat{h}(2\pi f)\vert^2{\rm d}f,
    \label{10}
\end{align}
where the second equality comes from Parseval's theorem, and $\hat{h}$ is the Fourier transform of $h$ \footnote{We clarify that in Eq.~(\ref{10}), $h(t)$ is actually taken as a step function equal to 0 in the interval $(-\infty,0)$ and given by Eq.~(\ref{9}) in the interval $[0,+\infty)$, otherwise the integral in Eq.~(\ref{10}) is divergent.}. We calculate this Fourier transform and take the one-sided energy spectrum to obtain ${\rm d}E_{\rm GW}/{\rm d}f$, which we plot in Fig.~(\ref{dEdf}) for $B_{\rm pole}=10^{15}$ G, a value within the range of the magnetic field strength at the surface of magnetars estimated from observations \cite{Olausen_2014}\footnote{Information on known galactic magnetars can be found at http://www.physics.mcgill.ca/~pulsar/magnetar/main.html.}. We get a maximum value of $\frac{{\rm d}E_{\rm GW}}{{\rm d}f}\big\vert_{f=f_{\rm fmode}}=2.5\times10^{37}{\rm erg}\hspace{1mm} {\rm s}$ and a total $E_{\rm GW}=5.0\times10^{37} {\rm erg}$.

Utilising fitting relations from alternative general relativistic magnetohydrodynamic simulations reported in \cite{2012_l} yields comparable energy estimates $E_{\rm GW}=4.7\times10^{37}$ erg. This value is slightly smaller than that obtained using the analytic expression from \cite{lasky2012gravitational}, implying our results throughout this paper are somewhat conservative with respect to the choice of these simulations. 
%\cite{lasky2012gravitational} is significantly more optimistic for higher values of $B_{pole}$, since it assumes a greater exponent $\alpha$ in the relation $E_{\rm GW}\propto (B_{\rm pole})^\alpha$.

\section{DETECTABILITY WITH 3G DETECTORS}\label{IV}
\begin{figure}
\centering
\includegraphics[scale=0.68]{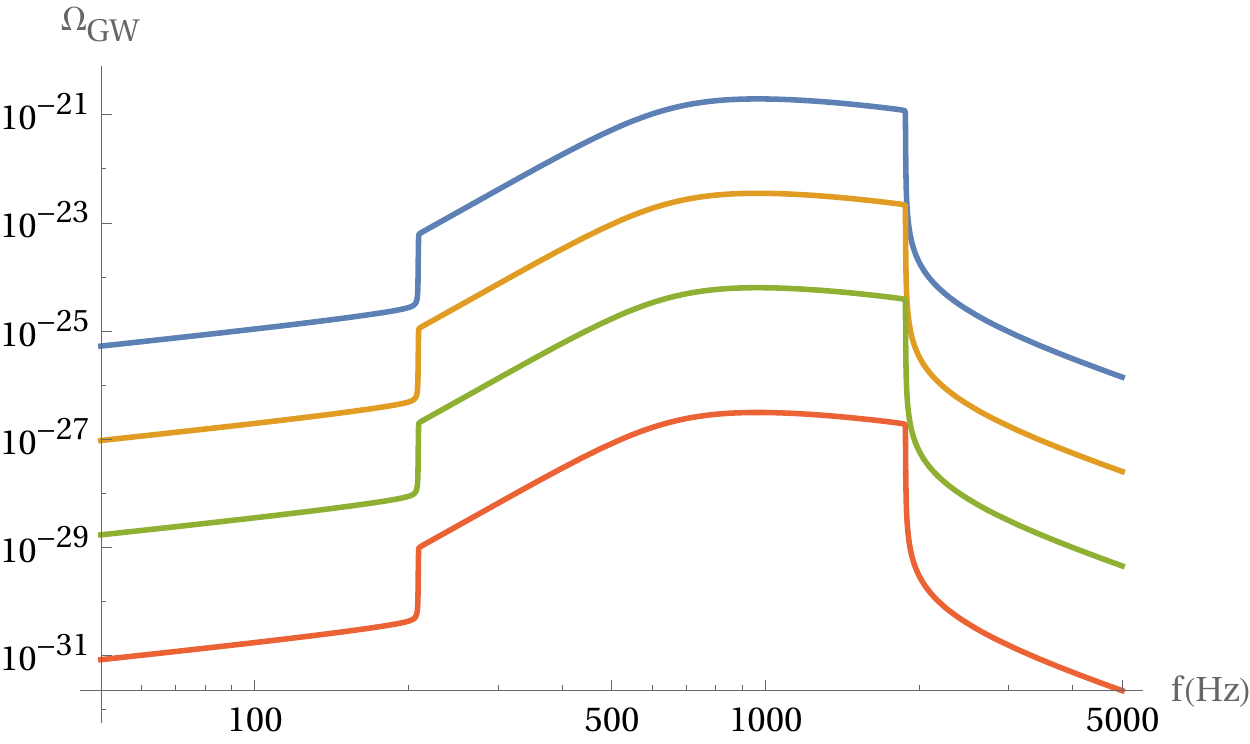}
\caption{The normalised GW energy spectrum of the magnetar GW background for $\lambda=0.02539M_{\odot}^{-1}$ and $B_{\rm pole}=10^{15}$, $5\times10^{14}$, $2.5\times10^{14}$, $10^{14}$ G (blue, yellow, green and red curve respectively). Cosmic expansion leads to the Doppler shifting of the $f_{\rm fmode}$ peak from $f=f_{\rm fmode}$ to $f=f_{\rm fmode}/z_{\rm max}$, in accordance with the SFR. The contributions for frequencies outside this range are significantly lower.}
\label{OmegaGW}
\end{figure}

Current estimates of the magnetic field strength of magnetars predict values $\sim(10^{14}-10^{15})$ G \cite{2011,2016,doi:10.1146/annurev-astro-081915-023329}. We plot  $\Omega_{\rm GW}$ for $B_{\rm pole}=10^{15}$ G, $5\times10^{14}$ G, $2.5\times10^{14}$ G, and $10^{14}$ G in Fig.~(\ref{OmegaGW}). Note the sudden drop in $\Omega_{\rm GW}$ below $f=f_{\rm fmode}/(1+z_{\rm max})=209.2$ Hz and beyond $f=f_{\rm fmode}=1883.1$ Hz. The total GW energy density $\Omega_{\rm GW}$ varies significantly with $B_{\rm pole}$. As one can see from Eqs.~(\ref{7})-(\ref{10}), a decrease of one order of magnitude in $B_{\rm pole}$ results in a decrease of almost six orders of magnitude in $\Omega_{\rm GW}$. We see that a value of $B_{\rm pole}=10^{15}$ G yields a maximum value $\Omega_{\rm GW}\sim10^{-21}$, which is an extremely small value compared to sensitivity estimates for third-generation gravitational-wave detectors.

The calculated signal is too weak to be detected by second-generation (2g) detectors. Even considering the combined aLIGO+aVirgo+KAGRA network, the threshold for detection is $\Omega_{\rm GW}\sim10^{-9}$  \cite{2021_s}. LISA's high sensitivity lies in a low frequency range up to $f=10^{-1}$ Hz \cite{2019}, far below the range where our signal peaks. Thus, we only consider 3g detectors, ET~\cite{EinsteinTelescope} (a triangle-shaped detector made up of three interferometers) and CE~\cite{CosmicExplorer} (an L-shaped interferometer). We compare the strength of our signal to a network of one ET and two CE detectors. The sensitivity curves for the networks of 2g and 3g detectors are obtained using power-law integration \cite{2013_tr}; see the Appendix for a detailed explanation on the example of ET+2CE.

\begin{figure}
\centering
\includegraphics[scale=0.68]{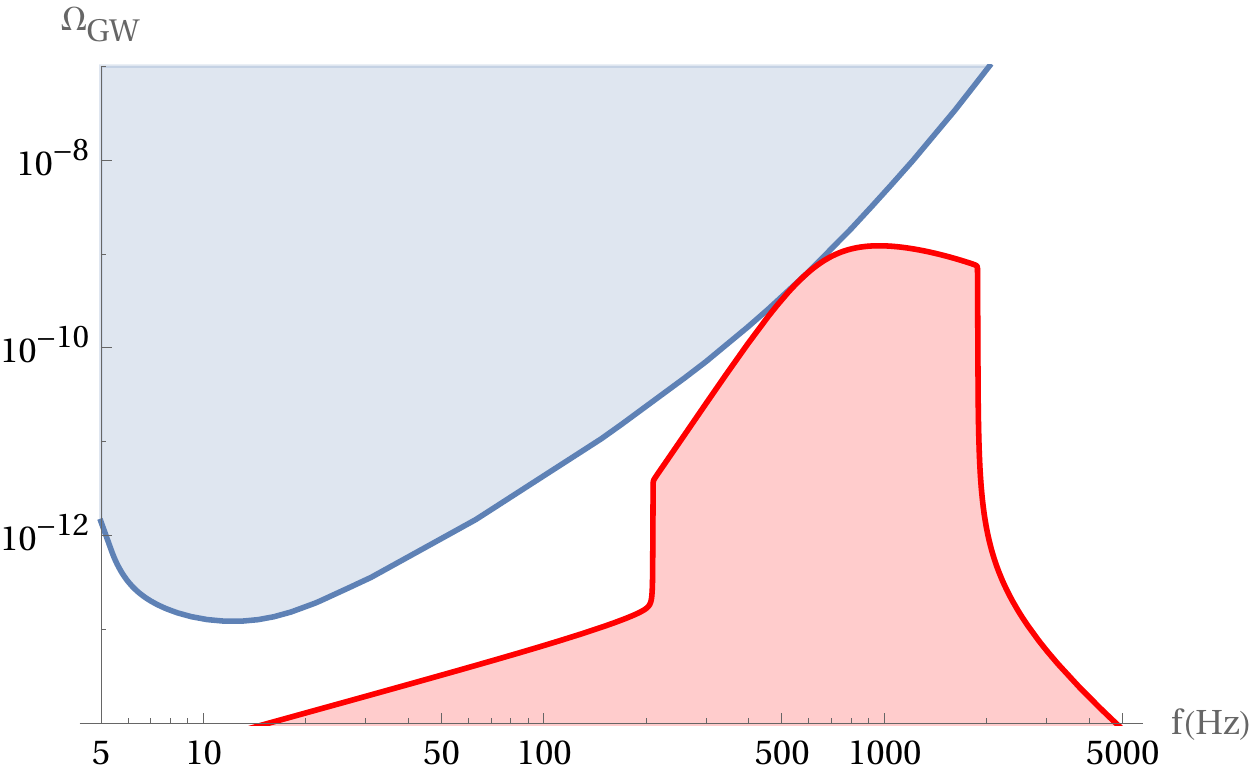}
\caption{Blue curve: $\Omega_{\rm GW}$ sensitivity of the ET$+2$CE network obtained via power-law integration for one year of coherent observations. Red curve: $\Omega_{\rm GW}$ of the magnetar GW background for $\lambda=1M_{\odot}^{-1}$ and $B_{\rm pole}=5.7\times10^{16}$ G. The blue and red shaded regions need to overlap so that the signal can be detected; in this case, it is barely detectable, at frequency $f=580$ Hz.}
\label{OmegaPI}
\end{figure}

Current estimates of $\sim(10^{14}-10^{15})$ G \cite{2011,2016,doi:10.1146/annurev-astro-081915-023329} for the magnetic field strength of magnetars are limited to their surface, while there is still a big uncertainty about the strength of the magnetic field in the magnetars' interior, which could reach values as high as  $(10^{16}-10^{17})$ G~\cite[e.g.,][]{2011_3,2014,2019_2}. For our reference value of $\lambda=0.025M_{\odot}^{-1}$, the minimum required $B_{\rm pole}$ for detection by the ET$+2$CE network is  $1.1\times 10^{17}$ G. For $B_{\rm pole}\sim 5\times10^{16}$ G, we obtain  $\lambda\geq2.2M_{\odot}^{-1}$. Assuming that the ratio of stars turning into magnetars is $\sim10^{-2}$ and that every magnetar emits  $\sim10^2$ MGFs in its lifetime, so that we obtain $\lambda=1M_{\odot}^{-1}$, leads to $B_{\rm pole}\geq5.7\times 10^{16}$ G. This is the best compromise of high values for $\lambda$ and $B_{\rm pole}$. Further reducing $B_{\rm pole}$ to $10^{16}G$ requires  $\lambda\geq2.5\times10^4M_{\odot}^{-1}$, an unrealistically high value. In all cases, minimum detection occurs at $f=580$ Hz (for our specific choice for the equation of state). We illustrate in Fig.~(\ref{OmegaPI}) the requirement for detection of the produced $\Omega_{\rm GW}$ from the ET$+2$CE network. It is clear that magnetic fields of $(10^{14}-10^{15}) $G that match estimates for the surface strength require completely unrealistic values for $\lambda$.

\section{CONCLUSIONS}\label{V}
We study the GW background resulting from giant magnetar flares throughout the Universe. Unlike previous studies focusing on the steady emission of GWs from non-zero stellar ellipticities, we explore the effect of MGFs on the GW background. We use  analytic expressions for the GW strain and energy obtained from general relativistic magnetohydrodynamic simulations of highly magnetized neutron stars. We consider a typical equation of state of $M=1.4M_{\odot}$ and $R=13$ km for the magnetar population. We also ignore the delay time between the star formation and the emission of MGFs, an approximation motivated by the short evolution time between star birth and supernovae as well as the young age of magnetars. There are large uncertainties in the values of the magnetic field at the pole $B_{\rm pole}$ and the scale factor between the MGF rate and the SFR $\lambda$. We explore these parameters in our analysis and find they significantly affect the estimate of the MGF GW background.

We conclude that the detection of the GW background from MGFs is unlikely with current or even next-generation gravitational-wave detectors. For the aforementioned choice of parameters, a number of MGFs per magnetar $\gtrsim10^2$ (so that $\lambda\gtrsim1M_\odot^{-1}$) and a magnetic field $B_{\rm pole}\gtrsim6\times10^{16}$ G are required for detection by the network of  one ET and two CE detectors. For lower values of $B_{\rm pole}$, the normalised energy spectrum falls many orders of magnitude, far lower than the sensitivity of any detector. Although we assume all magnetars to have the same properties (B field, mass, radius), taking a population distribution of these values is unlikely to significantly alter our qualitative conclusion that the background is not detectable. One could further consider other magnetic-field topologies, such as toroidal or twisted-torus magnetic fields, but we do not expect this to have a significant effect on our results. 
More data on the MGF rate and the strength of the magnetic field in the interior of magnetars would also prove insightful, although we are pessimistic that they would significantly alter our conclusions.

We stress that our conclusions are limited to the gravitational-wave background from MGFs. Other mechanisms may lead to a detectable gravitational-wave background from magnetars. For example, studies of GW emission due to non-zero stellar ellipticities from internal magnetic fields~\cite[e.g.,][]{refId0,2008_k,10.1111/j.1365-2966.2010.17861.x,PhysRevD.87.042002,universe7100381} estimate that the gravitational-wave background is detectable by 3g interferometers, although these results depend on a number of assumptions in the models that are currently not well understood. 

There is still much uncertainty about the GW energetics of MGFs, and there are a number of magnetars ($\sim$30 currently known \cite{Olausen_2014})
%\footnote{See http://www.physics.mcgill.ca/~pulsar/magnetar/main.html.}
 within/next to our galaxy. These remain a source of interest for gravitational-wave searches from individual bursts (such as \cite{2019_p}) with the current generation of gravitational-wave observatories.

\acknowledgements
N.K. is supported by King's College London through an NMES Funded Studentship.
P.D.L. is grateful to Teagan Clarke and Nikhil Sarin for early conversations about this work. P.D.L. is supported through Australian Research Council Future Fellowship FT160100112, Centre of Excellence CE170100004, and discovery Projects DP180103155 and DP220101610.
R.Q.J is supported by the U.S. National Science Foundation grants PHY-1921006 and PHY-2011334.
M.S. is supported in part by the Science and Technology Facility Council (STFC), United Kingdom, under the research grant ST/P000258/1. This paper has been given LIGO DCC number P2200061, and the Einstein Telscope ET-OSB number ET-0035A-22.

\section*{APPENDIX}
If there are enough GW events, the time between events will tend to be small compared to the duration of individual events and the overlapping signals will create a continuous background determined by the signals' spectral properties \cite{PhysRevD.87.042002}.
We can evaluate if MGFs are likely to form such a background by calculating the duty cycle \cite{2017_rom} (the ratio of the duration of a typical event to the mean time between events \cite{PhysRevD.87.042002}) describing a GW signal from an MGF:
\begin{equation}
    D=\int_0^{z_{\rm max}}R_{\rm MGF}(z)\tau(1+z){\rm d}z,
\end{equation}
where $1+z$ rescales $\tau$ to incorporate time dilation. For our conservative choice of $\lambda=0.025M_{\odot}^{-1}$, we find $D\simeq40\gg1$; the GW signals are expected to form a continuous GW background signal ideal for detection using cross-correlation.

To obtain the $\Omega_{\rm GW}$ sensitivity curve for the ET$+2$CE network, we perform a cross correlation of ET placed at the Virgo site and two CE detectors located at LIGO Hanford and Livingston. We calculate the variance for the cross correlation spectrum of two interferometers \cite{2021_2}
\begin{equation}
    \sigma_{\rm IJ}^2(f)\approx\frac{1}{2T{\rm \Delta} f}\frac{P_{\rm I}(f)P_{\rm J} (f)}{\gamma^2_{\rm IJ}(f)S^2_0(f)},
\end{equation}
where $T$ is the observation time, which we set at $1$ yr,   ${\rm \Delta} f=0.25$ Hz is the frequency resolution, $P_{\rm I}(f)$ is the one-sided power spectral density of the I-th detector and is simply obtained as the square of the effective strain noise $h_{\rm eff}$ of the detector (see \cite{2011_2} for ET and \cite{noauthor_ligo-t1500293-v13_nodate} for CE),  $\gamma_{\rm IJ}$ is the normalised overlap reduction function for the cross correlation of the I-th and J-th detector \cite{PhysRevD.59.102001} and $S_0(f)=(3H_0)^2/(10\pi^2f^3)$ \cite{mingarelli2019understanding}.

We calculate $\sigma_{\rm IJ(}f)$ for the ten different pairs resulting from the five interferometers (ET is made up of three interferometers and each CE consists of one). The total variance is
\begin{equation}
    \sigma_{\rm total}=\bigg(\sum_{\rm I}^5\sum_{\rm J>I}\frac{1}{\sigma_{\rm IJ}}\bigg)^{-1}.
\end{equation}
One can obtain a sensitivity curve from $\sigma_{\rm total} $ assuming that the normalised GW energy spectrum follows a power law \cite{2013_tr},
\begin{equation}
    \Omega_{\rm GW}(f)=\Omega_{\beta}\bigg(\frac{f}{f_{\rm ref}}\bigg)^\beta,
\end{equation}
and performing integration for a range of exponents $\beta$. We utilise a publicly available code for this process \cite{noauthor_datapicurvepy_nodate}, considering a reference frequency $f_{\rm ref}=25$ Hz.

%\end{multicols}

%\begin{multicols}{2}

\bibliographystyle{unsrt}
\bibliography{references}

%\end{multicols}

\end{document}